\documentclass[%
 reprint,
 amsmath,amssymb,
 aps,
]{revtex4-2}

\usepackage{comment}
\usepackage[normalem]{ulem}
\usepackage{graphicx}
\usepackage{dcolumn}
\usepackage{bm}
\usepackage{hyperref}

\usepackage{soul,xcolor}
\setstcolor{red}

\begin{document}

\title{Anomalous dynamical scaling determines universal critical singularities}

\author{Attilio L. Stella}
\affiliation{Department of Physics and Astronomy, University of Padova, Via Marzolo 8, I-35131 Padova, Italy}
\affiliation{INFN, Sezione di Padova, Via Marzolo 8, I-35131 Padova, Italy}
\author{Aleksei Chechkin}
\affiliation{Institute of Physics and Astronomy, University of Potsdam, D-14476 Potsdam-Golm, Germany,}
\affiliation{Faculty of Pure and Applied Mathematica, Hugo Steinhaus Center, University of Science and Technology, Wyspianskiego 27, 50-370 Wrocław, Poland,}
\affiliation{Akhiezer Institute for Theoretical Physics, 61108 Kharkov, Ukraine}
\author{Gianluca Teza}%
\email{gianluca.teza@weizmann.ac.il}
\affiliation{Department of Physics of Complex Systems, Weizmann Institute of Science, Rehovot 7610001, Israel}

\date{\today}

\begin{abstract}
Anomalous diffusion phenomena occur on length scales spanning from intracellular to astrophysical ranges. A specific form of decay at large argument of the probability density function of rescaled displacement (scaling function) is derived and shown to imply universal singularities in the normalized cumulant generator. Exact calculations for continuous time random walks provide paradigmatic examples connected with singularities of second order phase transitions. In the biased case scaling is restricted to displacements in the drift direction and singularities have no equilibrium analogue.
\end{abstract}

\maketitle

Scaling laws are at the basis of our understanding of equilibrium systems at criticality and of the singularities associated to second order phase transitions \cite{fisher1967theory,kadanoff2000statistical,stanley1999}.
A key role in this context is played by the scaling of probability density functions (PDF) of observables \cite{cardy2012}, like the spatial span of a self repelling polymer at varying backbone lengths \cite{fisher1959excluded,Fisher1966,kafri2002melting} or the magnetization of a finite Ising model for different system sizes \cite{binder1981,kaski1984}.

PDFs with an analogous type of scaling, but with time $t$ replacing chain length or system size, are also often met outside of equilibrium.
A paradigmatic example is that of anomalous spatial diffusion, where the mean squared displacement grows as $\left< x^2\right> \sim t^{2\nu}$ with $\nu\neq1/2$ ($\nu=1/2$ provides Brownian diffusion \cite{metzler2014anomalous}).
At long times, the associated PDF asymptotically satisfies
\begin{equation}\label{eq:scaling}
    p(x,t)\sim t^{-\nu} f(x/t^{\nu})
\end{equation}
where $f$ is a non-Gaussian scaling function \cite{bouchaud1990anomalous,cecconi2022probability}.
The importance of this characterization follows from the ubiquity of anomalous diffusion in nature, which can be observed in a variety of experiments carried out on different scales ranging from astrophysical to intracellular ones \cite{bouchaud1990anomalous,klages2008anomalous,metzler2000random,he2008random,sokolov2012models,metzler2014anomalous,golding2006physical,lubelski2008nonergodicity,weber2010bacterial,weigel2011ergodic,weber2012nonthermal,viswanathan1999optimizing,barthelemy2008levy,lagutin2003anomalous,uchaikin2013fractional}.
Most relevant is also the fact that for a whole class of diffusion problems, $f$ can be expected to decay exponentially fast at large values with a power of the argument linked to $\nu$ by a relation first established by Fisher for polymers in equilibrium \cite{Fisher1966},
and supported by probabilistic arguments and numerical model calculations \cite{havlin1987diffusion,cecconi2022probability,bouchaud1990anomalous}, simulations \cite{Havlin1985} and renormalization group results \cite{Guyer1985}.

Besides the polymer case, stretched exponential decays of scaling functions have been conjectured or numerically estimated also for equilibrium criticality, especially for the PDF of the magnetization of finite Ising systems at the Curie temperature and in zero magnetic field \cite{binder1981,kaski1984,mccoy2014two,eisenriegler1987,hilfer1994}. On the basis of these decays analogies between magnetic critical phenomena and anomalous diffusion were already stressed in early work \cite{bouchaud1990anomalous}.
In the magnetic Ising case the power quantifying the stretching is expected to be directly connected to the Kadanoff exponent determining the magnetic field singularity of the free energy density \cite{kadanoff1967static,kadanoff2000statistical}.
In the attempt to deepen its connection with singularities and other universal aspects of equilibrium criticality, this type of stretched exponential decay, but also modulated by a power law factor, was conjectured in Refs. \cite{Bruce1995,Hilfer1995}.
However, such conjecture could never be proven or fully confirmed numerically \cite{stauffer1998,tsypin2000probability,berg2002overlap,hilfer2003multicanonical,hilfer2005multicanonical}.

Progress in the understanding of non-equilibrium dynamics largely relies on parallels one can draw with equilibrium \cite{Touchette2009,touchette2013,teza2022eigenvalue}.
Thus, it is fundamental to investigate the possible connections established by the scaling function decays encountered in anomalous diffusion with the singularities observed in equilibrium systems at criticality.
A characterization in this perspective of scaling and its consequences for anomalous diffusion is also of interest for general non-equilibrium theory.

In this letter, we show that the scaling property of a diffusing system implies a specific form of decay of the scaling function, which determines power-law singularities in the scaled cumulant generating function (SCGF) of displacement.
Remarkably, we show that such singularities originate from the same form of decay of the scaling function once postulated for the magnetization of Ising systems, including the modulating power law factor \cite{Bruce1995,Hilfer1995}.
The singularities propagate to the large deviation functions \cite{Touchette2009,touchette2013} and cause divergences of a dynamical response function analogous to a magnetic susceptibility in equilibrium.
The Fisher relation is shown to follow from the property of extensivity in time of the generator of unnormalized cumulants.
We also address the case of biased diffusion, obtaining singularities corresponding to scaling forms that have no analogue in equilibrium systems.
All this is verified by exact calculations for continuous time random walks (CTRW) and related fractional drift-duffusion equations as generic models for anomalous diffusion \cite{metzler2000random,klages2008anomalous,metzler2014anomalous}.
In this way we establish bridges among three pillars of statistical mechanics: scaling, anomalous diffusion and large deviation theory.

We start by considering the general case of a particle diffusing on a 1-dimensional landscape, assuming that the scaling hypothesis of Eq. \ref{eq:scaling} holds for some exponent $0<\nu<1$.
The associated generating function is expressed in terms of a Laplace transform as $G(\lambda,t)=\int_{-\infty}^{+\infty} dx\ e^{\lambda x}p(x,t)$.
For long enough times, on the basis of Eq. \ref{eq:scaling}, it takes the form
\begin{equation}\label{eq:gener_scal}
    G(\lambda,t)\sim\int_{-\infty}^{+\infty} dz\  e^{\lambda t^{\nu} z} f(z)
\end{equation}
where we performed the variable change $z=x/t^{\nu}$.
The unnormalized cumulants are generated by differentiation with respect to $\lambda$ at $\lambda=0$ of $\log G$, which here we assume to scale asymptotically linearly in time.
Hence, the SCGF can be expressed through the limit
\begin{equation}\label{eq:SCGF}
    \varepsilon(\lambda)=\lim_{t\to\infty}\frac{1}{t}\log G(\lambda,t) .
\end{equation}
It is already apparent how the existence of this finite limit cannot exclude the possibility of a singularity of $\varepsilon(\lambda)$ at $\lambda=0$.
Indeed, the $n$-th order unnormalized cumulant grows as $t^{n\nu}$, implying that for a non-Gaussian scaling function all the scaled cumulants above a certain order become infinite.
Consistently, this can cause a divergence of the $n$-th derivative of the SCGF as soon as $n>1/\nu$.
In the case of free diffusion the scaling function $f(z)$ is symmetric like that of the magnetization of an Ising system at criticality. To the contrary, for biased diffusion we could expect an asymmetry or even a restriction of the domain in which Eq. \ref{eq:scaling} holds.

For both free and biased diffusion, the dominant behavior of the integral in Eq. \ref{eq:gener_scal} for large $\lambda t^{\nu}$ with $\lambda>0$ is determined by the decay of the scaling function $f(z)$ for $z\to +\infty$ (throughout this work when considering a bias we assume it to be in the positive direction).
Indeed, with $f$ differentiable and monotonically decreasing sufficiently fast to zero as $z \to +\infty$, the integrand in Eq. \ref{eq:gener_scal} reaches a maximum at some $\overline{z}$ which increases towards $+\infty$ as time grows.
Applying Laplace's method \cite{bender1978} the leading contribution to $G$ will come from the integrand in Eq.(2) computed at the value $\bar{z}$ satisfying $f'(\bar{z})/f(\bar{z})=-\lambda t^{\nu}$, which maximizes the argument of the exponential in terms of which one can write the integrand in Eq. \ref{eq:gener_scal}.
Assuming that $\bar{z}$ for long enough times grows as $(\lambda t^{\nu})^{1/\delta}$ for some $\delta>0$, we get that the differential equation $f'(\bar{z})/f(\bar{z})\sim-\bar{z}^\delta$ asymptotically holds.
It admits the solution
\begin{equation}\label{eq:scal_func_sol}
f(\overline{z}) \sim  \bar{z}^{\psi}e^{- c \bar{z}^{\delta +1}}
\end{equation}
for some positive constant $c$ and any exponent $\psi$. The factor $\bar{z}^{\psi}$ is introduced to allow the possibility of cancellation of a term $\propto \log(\lambda t^\nu
)$ in $\log G$, as shown below, and implies a correction $\propto \bar{z}^{-1}$ to the differential equation.

The exponent $\delta$ not only enters the tails of the scaling function, but also determines the asymptotic dominant term in $\varepsilon$.
Indeed, substituting $\bar{z}$ in Eq. \ref{eq:gener_scal} and using Laplace's method we get
\begin{eqnarray}\label{eq:logG_delta_nu}
    \log G(\lambda,t)\ && \sim \lambda t^{\nu}\bar{z}-c \bar{z}^{\delta +1} + \\ \nonumber
    && +\frac{2\psi+1-\delta}{2} \log \bar{z} + \text{const.}
\end{eqnarray}
up to a correction $\propto \bar{z}^{-1-\delta}$ \cite{SM}.
Recalling that $\bar{z}^\delta\sim \lambda t^{\nu}$, we find that the first two terms are proportional to the same powers of $t$ and of $\lambda$, the third term is a logarithmic correction (disappearing only for $\psi=(\delta-1)/2$)  and the fourth is a time independent constant \cite{SM}.
The term $\propto \log(\overline{z})$ is the only one which actually allows to split the $\lambda$ and $t$ dependencies into the sum of two separate terms. Therefore its presence would introduce a logarithmic singular dependence on $\lambda$ in the whole $t$-independent part of $\log G$. For such reason, we expect that this dependence should be dropped by the above choice of $\psi$. 

A contribution extensive in $t$ for $\log G$ can result from the first two terms. Since the exponent of $t$ depends  on both $\nu $ and $\delta$, Eq. \ref{eq:SCGF} provides $\delta=\nu/(1-\nu)$, known as the Fisher relation \cite{Fisher1966}.
We also find that Eq. \ref{eq:logG_delta_nu} predicts a singular dependence for $\varepsilon(\lambda) \sim \lambda^{1/\nu}$ at $\lambda =0^+$.
Therefore, the larger is $\delta$, i.e. the faster the decay of $f$, the larger also $\nu$ and thus the stronger the singularity in $\lambda$. The fact that it is determined by the asymptotic rate of decay of $f$ confers a universal character to the singularity: different $f$'s can have the same law of decay and thus cause the same singularity. 

Below we demonstrate the existence of a critical singularity in the SCGF of the CTRW in both free (subdiffusive) and biased (subdiffusive and superdiffusive) regimes \cite{montroll1965random,kenkre1973generalized,SM}.
In this model, a particle jumps on a one-dimensional lattice with spacing $L$ with a rate $r$ ($l$) to the right (left) nearest neighboring site.
The waiting times $\tau$ in-between the jumps occur according to a certain PDF $\omega(\tau)$.
Anomalous diffusion occurs when this PDF decays to zero as a power law ( $\omega(\tau)\sim \tau^{-1-\alpha}$ with $0<\alpha<1$) for $\tau \to \infty$, so that its first moment is infinite.
The probability $P_i(t)$ to observe the particle on the $i$-th site at a certain time $t$ evolves according to the generalized master equation
\begin{equation}\label{eq:ME_CTRW}
    \partial^{\alpha}_{t}P_i(t)= r P_{i-1}(t) + l P_{i+1}(t)-(r+l)P_i(t)
\end{equation}
where $\partial^{\alpha}_{t}$ is the $\alpha$-order Caputo fractional derivative (implying that the unit of $r$ and $l$ is $[\textrm{time}]^{-\alpha}$) which has the integral representation \cite{carpinteri2014fractals,SM}
\begin{equation}\label{eq:caputo}
    \partial_t^\alpha P_i(t)=\frac{1}{\Gamma(1-\alpha)}~\int_0^t \frac{\partial_\tau P_i(\tau)}{(t-\tau)^{\alpha}} d\tau
\end{equation}
where $\Gamma(\cdot)$ is the complete Gamma function.
With initial condition $P_i(0)=\delta_{i,0}$, the generating function $G(\lambda,t)=\Sigma_i e^{\lambda L i} P_i(t)$
satisfies
\begin{equation}
    [\partial^{\alpha}_{t}-\varepsilon_B(\lambda)]G(\lambda,t) =0
\end{equation}
where $\varepsilon_B(\lambda)=r(e^{L\lambda}-1)+l(e^{-L\lambda}-1)$ is the SCGF of Brownian ($\alpha=1$) diffusion \cite{teza2020exact,teza2020thesis}.
The solution can be found passing to Laplace space \cite{SM} to be
\begin{equation}\label{eq:gen_CTRW}
    G(\lambda,t)=E_{\alpha}(\varepsilon_B(\lambda)t^{\alpha})
\end{equation} 
where $E_{\alpha}$ is the one-parameter Mittag-Leffler function \cite{gorenflo2020mittag}.
The asymptotics of this function are proportional to $e^{\varepsilon_B(\lambda)^{1/\alpha}t}$ and $-1/\varepsilon_B(\lambda)t^{\alpha}$ for positive and negative arguments, respectively.

Let us first address the unbiased case, with $r=l$. We find that $\varepsilon_B(\lambda)=(4r)\sinh^2 (L\lambda/2)\geq0$. Therefore, taking the long time limit provides us with the following SCGF
\begin{equation}\label{eq:SCGF_unbiased_CTRW}
    \varepsilon(\lambda)=(4r)^{1/\alpha}\sinh^{2/\alpha} (L\lambda/2)\sim \lambda^{2/\alpha} +O(\lambda^{2+2/\alpha})
\end{equation}
Thus we get a leading singularity $\sim \lambda^{2/\alpha}$  in the scaled cumulant generating function at $\lambda=0$. This singularity is qualitatively of the same type encountered for the free energy density of Ising systems at criticality, with $\lambda$ here playing the role of magnetic field there \cite{mccoy2014two}. The first derivative of $\varepsilon(\lambda)$ divergent at $\lambda = 0$ is that of even order $n$ with $n$ just exceeding $2/\alpha$. 
This derivative assumes the meaning of a diverging dynamical response function analogous to the magnetic susceptibility of an Ising model at criticality.

The correctness of our general argument connecting the singularity of the SCGF to a specific form of asymptotic decay of the scaling function can be exactly verified in this example.
Indeed, in the case $r=l$ the continuum limit of Eq. \ref{eq:ME_CTRW} yields the fractional diffusion equation \cite{metzler2000random,SM} regulating the PDF $P_i(t)/L\to p(x,t)$ of observing the particle at position $iL\to x$ at time $t$
\begin{equation}\label{eq:FD_eq}
    \partial^{\alpha}_t p(x,t)= D \partial^2_x p(x,t)\ ,
\end{equation}
where $r L^2 \to D$ is a diffusion constant, which to the purpose of our further discussion can be assumed equal to $1$.
The solution to this equation satisfies exactly the scaling form of Eq. \ref{eq:scaling} for all $x$ and $t$ with $\nu=\alpha/2$.
The scaling function can be expressed via the Fox H-function \cite{schneider1989fractional}, M-Wright function \cite{mainardi1994special}, or the one-sided L\'evy stable density \cite{barkai2000continuous}.
Choosing the M-Wright function, we can express the generating function as in Eq. \ref{eq:gener_scal} with $M_{\alpha/2}(z)$ replacing $f(z)$.
The tails of this scaling function are given by \cite{mainardi1994special,mainardi2010wright}
\begin{equation}\label{eq:m_wright}
    M_{\nu}(z) \sim |z|^{\frac{\nu-1/2}{1-\nu}} e^{-\frac{1-\nu}{\nu}|\nu z|^{1/(1-\nu)}}
\end{equation}
and have precisely the general form argued in Eq. \ref{eq:scal_func_sol}.
The exponent $\delta$ is found to take the value $\nu/(1-\nu)$ consistent with the Fisher relation and implying that $\log G$ grows linearly with time (from Eq. \ref{eq:logG_delta_nu}).
The multiplicative power factor with $\psi=(\nu -1/2)/(1-\nu)$, which translates in $\psi=(\delta -1)/2$, allows to drop the possible logarithmic dependence  on $\bar{z}$  for $\log G(\lambda,t)$, and the leading singularity is $\varepsilon(\lambda)\sim \lambda^{2/\alpha}$, confirming the result in Eq. \ref{eq:SCGF_unbiased_CTRW}. 

\begin{figure}
    \centering
    \includegraphics[width=\linewidth]{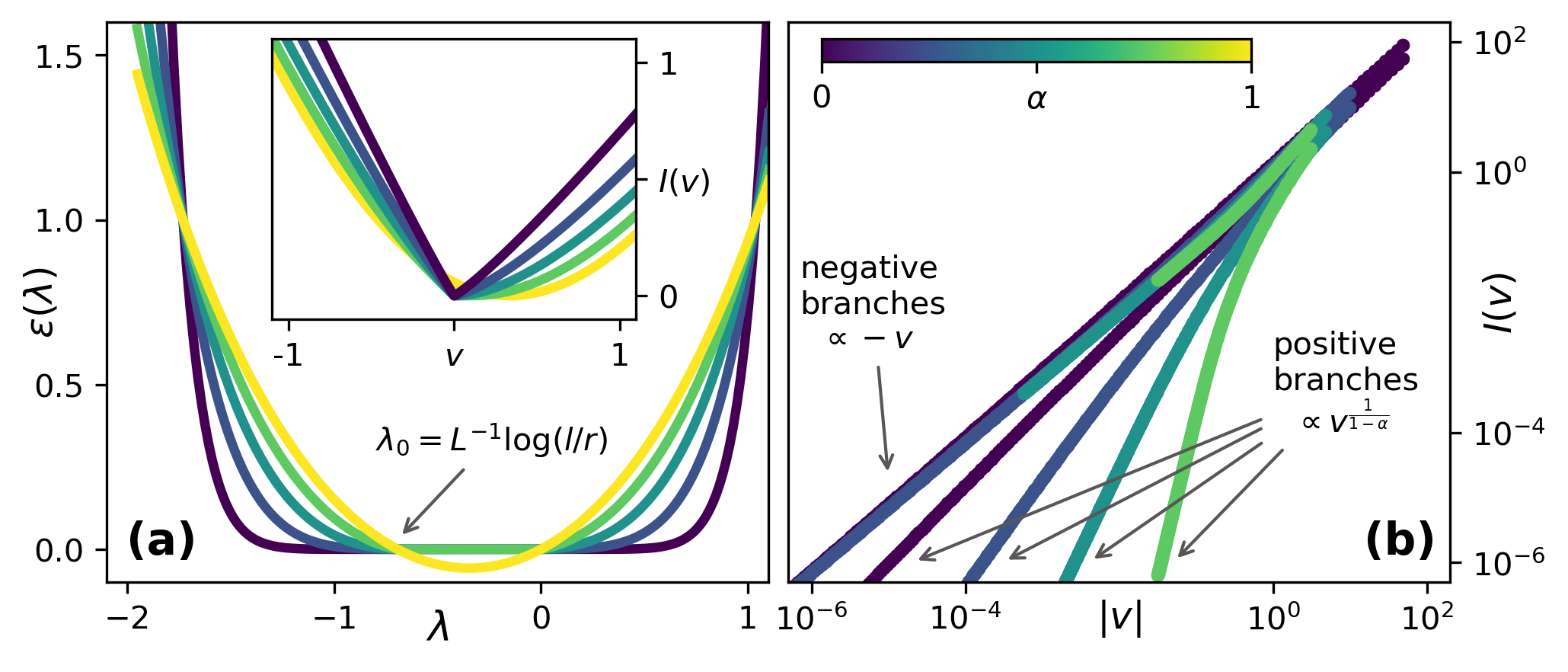}
    \caption{(a) SCGF for a biased ($r=2/3$) subdiffusive random walk (Eq. \ref{eq:SCGF_biased_CTRW}) for different values of $\alpha$: $\varepsilon(\lambda)$ identically zero for $\lambda_0\leq\lambda\leq0$.
    The non-singular case of Brownian diffusion ($\varepsilon_B$) is also reported for reference (yellow).
    (b) The branches of the rate functions exhibit a dependence $\sim v^{1/(1-\alpha)}$ for positive values, while $\sim -v$ for negative values, which ensures validity of the fluctuation theorem.
    }
    \label{fig:biased}
\end{figure}

Let us now address the biased CTRW case, given by Eq. \ref{eq:ME_CTRW} when $r\neq l$.
In such scenario, the factor $\varepsilon_B(\lambda)$ determining the lattice generating function $G(\lambda,t)$ has an additional zero $\lambda_0=L^{-1}\log l/r<0$ for $r>l$, as shown by the yellow curve in Fig. \ref{fig:biased}a.
This implies for $G(\lambda,t)$ a power-law asymptotic behavior $\propto - 1/\varepsilon_B(\lambda)t^{\alpha}$ in the infinite time limit for $\lambda_0<\lambda<0$, while an exponential dependence $\propto e^{\varepsilon_B(\lambda)^{1/\alpha}t}$ holds elsewhere.
This causes the associated SCGF to be identically zero in that interval, giving us
\begin{equation}\label{eq:SCGF_biased_CTRW}
\varepsilon(\lambda)=
    \begin{cases}
    \varepsilon_B(\lambda)^{1/\alpha} & \lambda \leq \lambda_0 \text{ and } \lambda \geq 0 \\
    0 & \lambda_0<\lambda<0
    \end{cases}
\end{equation}
which shows a power law singularity $\sim\lambda^{1/\alpha}$ for $\lambda \to 0^{+}$ (see Fig. \ref{fig:biased}a). An additional singularity $\sim (\lambda_0 -\lambda)^{1/\alpha}$ appears for $\lambda \to \lambda_0^{-}$. The simultaneous presence of these singularities is consistent with the fact that the SCGF satisfies the Gallavotti Cohen identity, $\varepsilon(\lambda)= \varepsilon(-\lambda +\lambda_0)$, making the function symmetric with respect to the $\lambda_0/2$ axis and heralding validity of the fluctuation theorem. 
Both these singularities are totally asymmetric, since $\varepsilon(\lambda)$ is identically zero for $\lambda_0<\lambda<0$. This asymmetry and the simultaneous presence escape analogies with equilibrium criticality. Nevertheless, the singularities can still be justified along the lines proposed here, but in terms of a scaling which now holds only in the positive $x$ domain.

The continuum limit of Eq. \ref{eq:ME_CTRW} for $r>l$ leads to the fractional drift diffusion equation \cite{SM}
\begin{equation}\label{eq:FDDE}
    \partial^{\alpha}_t p(x,t)= [-K\partial_x +D \partial^2_x ]p(x,t)\ ,
\end{equation}
where $(r-l)L\to K$ defines the drift constant and $(r+l)L^2/2 \to D$ the diffusion constant, implying that Eq. \ref{eq:FD_eq} is immediately recovered in the case $r=l$.
As in the free case, these limit prescriptions are justified by their consistency with the scaling of the solution of the resulting continuum equation.
Local detailed balance \cite{katz1983phase,teza2020rate,teza2020thesis} allows to link the coefficients to the rates of the CTRW \cite{chechkin2009fluctuation,SM}.
Since also in this case the actual values of the constants do not affect our results, for simplicity we set $K=1$ and $D=1$ below.  
An asymptotic ($t\to\infty$) solution of Eq. \ref{eq:FDDE} for the positive branch ($x>0$) is found to be $t^{-\alpha} M_\alpha(x/t^\alpha)$\cite{chechkin2009fluctuation,sokolov2009universal,mainardi2010wright,dieterich2015fluctuation}, with the M-Wright function playing again the role of scaling function. Indirect evidence of such scaling comes also from renormalization group calculations \cite{teza2022renormalization}.

The singularity implied by Eq. \ref{eq:SCGF_biased_CTRW} at $\lambda=0^+$ can again be directly obtained by our asymptotic analysis, since the behavior of $M_\nu(z)$ given in Eq. \ref{eq:m_wright} at large positive $z$ holds in the whole $0<\nu<1$ interval and is of the form proposed in Eq. \ref{eq:scal_func_sol}, now with $z=x/t^\alpha$. Thus, our derivation shows that also in this biased case the scaling function at large positive $z$ is consistent with $\varepsilon(\lambda)\sim \lambda^{1/\alpha}$ for $\lambda \to 0^+$ and with the Fisher relation, for both sub-diffusion ($\nu=\alpha <1/2$) and super-diffusion ($\nu=\alpha>1/2$).

A remarkable feature of the biased case is that the solution of Eq. \ref{eq:FDDE} is not satisfying scaling in the $t\to\infty$ limit for $x<0$. Indeed, for $x<0$ one finds 
$p(x,t) \sim t^{-\alpha} e^x$ \cite{SM}, which also shows that the total probability of negative displacements tends to annihilate in the limit. However, the global behavior of $\varepsilon(\lambda)$ can be exactly connected with the solution $p(x,t)$ of
Eq.\ref {eq:FDDE} on the whole $x$ axis by switching to Laplace transform in time for both $p$ and $G$ \cite{SM} and exploiting results in \cite{chechkin2009fluctuation}.
The biased case provides a remarkable example of the consequences for critical singularities of a probability distribution exhibiting one-sided scaling, outlining a scenario with no counterpart in equilibrium.

Integrating our results with the framework of large deviation theory allows to characterize direct implications on fluctuations \cite{Touchette2009}.
Since $\varepsilon(\lambda)$ is convex and differentiable, for the PDF of the "velocity" variable $v=x/t$ a rate function $I(v)$, satisfying for $t\to\infty$ 
\begin{equation}\label{eq:LDP}
    p(x/t=v,t)\sim e^{-t I\left(v\right)} \ ,
\end{equation}
can be obtained by application of the G\"{a}rtner-Ellis theorem \cite{gartner1977on,ellis1984large} as Legendre-Fenchel transform of $\varepsilon$ \cite{rockafellar1970convex}:
\begin{equation}\label{eq:rate_func_legendre}
    I(v)=\sup_{\lambda\in\mathbb{R}}\left[ v\lambda-\varepsilon(\lambda) \right]
\end{equation}  
This rate function is equal to zero at $v=0$, consistently with the fact that there is no conventional current in biased anomalous diffusion. Simple calculations \cite{SM} also show that the branches of $\varepsilon$ at $\lambda \sim 0^+$ and $\lambda \sim\lambda_0^-$ determine, respectively, $I(v) \sim v^{1/(1-\alpha)}$ around $v=0^+$ and $I(v)\sim -v$ around $v=0^-$. So, the two singularities of $\varepsilon$ merge into a single singular point at $v=0$, where $I(v)$ is clearly asymmetric.   

The Gallavotti-Cohen symmetry of the SCGF in Eq. \ref{eq:SCGF_biased_CTRW} allows also to show
that $I(v)-I(-v)=\lambda_0 v$ for any $v$ \cite{SM}, confirming validity of the fluctuation theorem \cite{lebowitz1999gallavotti,teza2020exact,teza2020thesis} for biased
anomalous diffusion, consistently with earlier results in Ref. \cite{chechkin2009fluctuation}.
Remarkably, the separate analysis of the branches of the rate function clearly shows that only the negative branch -- the one contrary to the bias -- is providing a linear contribution in $v$, which ultimately ensures validity of the fluctuation theorem.

The examples discussed above are of anomalous diffusion satisfying the Fisher relation $\delta=\nu/(1-\nu)$, even in the biased case. This relation, which we could clearly link to the requirement of standard extensivity in time for the unnormalized cumulant generator, is expected to be satisfied by a large class of problems in which the diffusion step does not depend on position in space \cite{cecconi2022probability}.
This class includes diffusion on fractals \cite{Guyer1985} and on percolation clusters \cite{Havlin1985}.
Still, if one gives up the requirement of standard extensivity for $\log G$ our approach applies also to problems belonging to the Richardson class \cite{richardson1926atmospheric,boffetta2002relative,cecconi2022probability} -- related to processes with position-dependent step sizes -- by enforcing the characterizing relation $\delta=(1-\nu)/\nu$. For this class our Eq. \ref{eq:logG_delta_nu} foresees a non standard extensivity in time, i.e. $\log G \propto t^{\nu/(1-\nu)}$,
and a singularity $\propto \lambda^{1/(1-\nu)}$ for the consistently defined $\varepsilon$.
We could verify these properties by exact calculations for a process of anomalous diffusion in inhomogeneous medium introduced in
\cite{cherstvy2013,stella2022inpreparation}.
Thus, our approach enables to characterize anomalous diffusion belonging to both Fisher and Richardson classes.

In summary we demonstrated the existence of power law critical singularities for the SCGF's of paradigmatic models of anomalous diffusion. Exact results fully support the link we predict between these singularities and the decay of the non-Gaussian scaling function of the displacement at asymptotic absolute values of its argument. The Fisher relation linking this decay to the diffusion exponent is shown to follow from the extensivity in time of the unnormalized cumulant generator, while a peculiar power law pre-factor excludes corrections $\propto \log(t)$ for this function. For biased diffusion two singularities with extreme asymmetry simultaneously result from a PDF with one-sided scaling, determining peculiar singular behavior in the rate function,
but not preventing validity of the FT.
We have also shown that the singularity generation mechanism valid for anomalous diffusion is based on exactly the same form of scaling function decay once postulated, but never proved, for Ising systems in equilibrium \cite{Bruce1995,Hilfer1995}. This suggests the possibility of a general probabilistic explanation of this form in anomalous scaling.

\begin{acknowledgments}
G. T. is supported by the Center for Statistical Mechanics at the Weizmann Institute of Science, the grant 662962 of the Simons foundation, the grants HALT and Hydrotronics of the EU Horizon 2020 program and the NSF-BSF grant 2020765. A. C. acknowledges support of the Polish National Agency for Academic Exchange (NAWA).
We thank David Mukamel and Oren Raz for useful discussions.
\end{acknowledgments}


\bibliography{refs.bib}

\providecommand{\noopsort}[1]{}\providecommand{\singleletter}[1]{#1}%
\begin{thebibliography}{68}%
\makeatletter
\providecommand \@ifxundefined [1]{%
 \@ifx{#1\undefined}
}%
\providecommand \@ifnum [1]{%
 \ifnum #1\expandafter \@firstoftwo
 \else \expandafter \@secondoftwo
 \fi
}%
\providecommand \@ifx [1]{%
 \ifx #1\expandafter \@firstoftwo
 \else \expandafter \@secondoftwo
 \fi
}%
\providecommand \natexlab [1]{#1}%
\providecommand \enquote  [1]{``#1''}%
\providecommand \bibnamefont  [1]{#1}%
\providecommand \bibfnamefont [1]{#1}%
\providecommand \citenamefont [1]{#1}%
\providecommand \href@noop [0]{\@secondoftwo}%
\providecommand \href [0]{\begingroup \@sanitize@url \@href}%
\providecommand \@href[1]{\@@startlink{#1}\@@href}%
\providecommand \@@href[1]{\endgroup#1\@@endlink}%
\providecommand \@sanitize@url [0]{\catcode `\\12\catcode `\$12\catcode
  `\&12\catcode `\#12\catcode `\^12\catcode `\_12\catcode `\%12\relax}%
\providecommand \@@startlink[1]{}%
\providecommand \@@endlink[0]{}%
\providecommand \url  [0]{\begingroup\@sanitize@url \@url }%
\providecommand \@url [1]{\endgroup\@href {#1}{\urlprefix }}%
\providecommand \urlprefix  [0]{URL }%
\providecommand \Eprint [0]{\href }%
\providecommand \doibase [0]{https://doi.org/}%
\providecommand \selectlanguage [0]{\@gobble}%
\providecommand \bibinfo  [0]{\@secondoftwo}%
\providecommand \bibfield  [0]{\@secondoftwo}%
\providecommand \translation [1]{[#1]}%
\providecommand \BibitemOpen [0]{}%
\providecommand \bibitemStop [0]{}%
\providecommand \bibitemNoStop [0]{.\EOS\space}%
\providecommand \EOS [0]{\spacefactor3000\relax}%
\providecommand \BibitemShut  [1]{\csname bibitem#1\endcsname}%
\let\auto@bib@innerbib\@empty
\bibitem [{\citenamefont {Fisher}(1967)}]{fisher1967theory}%
  \BibitemOpen
  \bibfield  {author} {\bibinfo {author} {\bibfnamefont {M.~E.}\ \bibnamefont
  {Fisher}},\ }\bibfield  {title} {\bibinfo {title} {The theory of equilibrium
  critical phenomena},\ }\href {https://doi.org/10.1088/0034-4885/30/2/306}
  {\bibfield  {journal} {\bibinfo  {journal} {Reports on progress in physics}\
  }\textbf {\bibinfo {volume} {30}},\ \bibinfo {pages} {615} (\bibinfo {year}
  {1967})}\BibitemShut {NoStop}%
\bibitem [{\citenamefont {Kadanoff}(2000)}]{kadanoff2000statistical}%
  \BibitemOpen
  \bibfield  {author} {\bibinfo {author} {\bibfnamefont {L.~P.}\ \bibnamefont
  {Kadanoff}},\ }\href@noop {} {\emph {\bibinfo {title} {Statistical physics:
  statics, dynamics and renormalization}}}\ (\bibinfo  {publisher} {World
  Scientific},\ \bibinfo {year} {2000})\BibitemShut {NoStop}%
\bibitem [{\citenamefont {Stanley}(1999)}]{stanley1999}%
  \BibitemOpen
  \bibfield  {author} {\bibinfo {author} {\bibfnamefont {H.~E.}\ \bibnamefont
  {Stanley}},\ }\bibfield  {title} {\bibinfo {title} {Scaling, universality,
  and renormalization: Three pillars of modern critical phenomena},\
  }\href@noop {} {\bibfield  {journal} {\bibinfo  {journal} {Reviews of modern
  physics}\ }\textbf {\bibinfo {volume} {71}},\ \bibinfo {pages} {S358}
  (\bibinfo {year} {1999})}\BibitemShut {NoStop}%
\bibitem [{\citenamefont {Cardy}(2012)}]{cardy2012}%
  \BibitemOpen
  \bibfield  {author} {\bibinfo {author} {\bibfnamefont {J.}~\bibnamefont
  {Cardy}},\ }\href@noop {} {\emph {\bibinfo {title} {Finite-size scaling}}}\
  (\bibinfo  {publisher} {Elsevier},\ \bibinfo {year} {2012})\BibitemShut
  {NoStop}%
\bibitem [{\citenamefont {Fisher}\ and\ \citenamefont
  {Sykes}(1959)}]{fisher1959excluded}%
  \BibitemOpen
  \bibfield  {author} {\bibinfo {author} {\bibfnamefont {M.~E.}\ \bibnamefont
  {Fisher}}\ and\ \bibinfo {author} {\bibfnamefont {M.~F.}\ \bibnamefont
  {Sykes}},\ }\bibfield  {title} {\bibinfo {title} {Excluded-volume problem and
  the ising model of ferromagnetism},\ }\href
  {https://doi.org/10.1103/PhysRev.114.45} {\bibfield  {journal} {\bibinfo
  {journal} {Phys. Rev.}\ }\textbf {\bibinfo {volume} {114}},\ \bibinfo {pages}
  {45} (\bibinfo {year} {1959})}\BibitemShut {NoStop}%
\bibitem [{\citenamefont {Fisher}(1966)}]{Fisher1966}%
  \BibitemOpen
  \bibfield  {author} {\bibinfo {author} {\bibfnamefont {M.~E.}\ \bibnamefont
  {Fisher}},\ }\bibfield  {title} {\bibinfo {title} {Shape of a self‐avoiding
  walk or polymer chain},\ }\href {https://doi.org/10.1063/1.1726734}
  {\bibfield  {journal} {\bibinfo  {journal} {The Journal of Chemical Physics}\
  }\textbf {\bibinfo {volume} {44}},\ \bibinfo {pages} {616} (\bibinfo {year}
  {1966})}\BibitemShut {NoStop}%
\bibitem [{\citenamefont {Kafri}\ \emph {et~al.}(2002)\citenamefont {Kafri},
  \citenamefont {Mukamel},\ and\ \citenamefont {Peliti}}]{kafri2002melting}%
  \BibitemOpen
  \bibfield  {author} {\bibinfo {author} {\bibfnamefont {Y.}~\bibnamefont
  {Kafri}}, \bibinfo {author} {\bibfnamefont {D.}~\bibnamefont {Mukamel}},\
  and\ \bibinfo {author} {\bibfnamefont {L.}~\bibnamefont {Peliti}},\
  }\bibfield  {title} {\bibinfo {title} {Melting and unzipping of dna},\ }\href
  {https://doi.org/10.1140/epjb/e20020138} {\bibfield  {journal} {\bibinfo
  {journal} {The European Physical Journal B-Condensed Matter and Complex
  Systems}\ }\textbf {\bibinfo {volume} {27}},\ \bibinfo {pages} {135}
  (\bibinfo {year} {2002})}\BibitemShut {NoStop}%
\bibitem [{\citenamefont {Binder}(1981)}]{binder1981}%
  \BibitemOpen
  \bibfield  {author} {\bibinfo {author} {\bibfnamefont {K.}~\bibnamefont
  {Binder}},\ }\bibfield  {title} {\bibinfo {title} {Finite size scaling
  analysis of ising model block distribution functions},\ }\href@noop {}
  {\bibfield  {journal} {\bibinfo  {journal} {Zeitschrift f{\"u}r Physik B
  Condensed Matter}\ }\textbf {\bibinfo {volume} {43}},\ \bibinfo {pages} {119}
  (\bibinfo {year} {1981})}\BibitemShut {NoStop}%
\bibitem [{\citenamefont {Kaski}\ \emph {et~al.}(1984)\citenamefont {Kaski},
  \citenamefont {Binder},\ and\ \citenamefont {Gunton}}]{kaski1984}%
  \BibitemOpen
  \bibfield  {author} {\bibinfo {author} {\bibfnamefont {K.}~\bibnamefont
  {Kaski}}, \bibinfo {author} {\bibfnamefont {K.}~\bibnamefont {Binder}},\ and\
  \bibinfo {author} {\bibfnamefont {J.}~\bibnamefont {Gunton}},\ }\bibfield
  {title} {\bibinfo {title} {Study of cell distribution functions of the
  three-dimensional ising model},\ }\href@noop {} {\bibfield  {journal}
  {\bibinfo  {journal} {Physical Review B}\ }\textbf {\bibinfo {volume} {29}},\
  \bibinfo {pages} {3996} (\bibinfo {year} {1984})}\BibitemShut {NoStop}%
\bibitem [{\citenamefont {Metzler}\ \emph {et~al.}(2014)\citenamefont
  {Metzler}, \citenamefont {Jeon}, \citenamefont {Cherstvy},\ and\
  \citenamefont {Barkai}}]{metzler2014anomalous}%
  \BibitemOpen
  \bibfield  {author} {\bibinfo {author} {\bibfnamefont {R.}~\bibnamefont
  {Metzler}}, \bibinfo {author} {\bibfnamefont {J.-H.}\ \bibnamefont {Jeon}},
  \bibinfo {author} {\bibfnamefont {A.~G.}\ \bibnamefont {Cherstvy}},\ and\
  \bibinfo {author} {\bibfnamefont {E.}~\bibnamefont {Barkai}},\ }\bibfield
  {title} {\bibinfo {title} {Anomalous diffusion models and their properties:
  non-stationarity, non-ergodicity, and ageing at the centenary of single
  particle tracking},\ }\href {https://doi.org/10.1039/C4CP03465A} {\bibfield
  {journal} {\bibinfo  {journal} {Physical Chemistry Chemical Physics}\
  }\textbf {\bibinfo {volume} {16}},\ \bibinfo {pages} {24128} (\bibinfo {year}
  {2014})}\BibitemShut {NoStop}%
\bibitem [{\citenamefont {Bouchaud}\ and\ \citenamefont
  {Georges}(1990)}]{bouchaud1990anomalous}%
  \BibitemOpen
  \bibfield  {author} {\bibinfo {author} {\bibfnamefont {J.-P.}\ \bibnamefont
  {Bouchaud}}\ and\ \bibinfo {author} {\bibfnamefont {A.}~\bibnamefont
  {Georges}},\ }\bibfield  {title} {\bibinfo {title} {Anomalous diffusion in
  disordered media: statistical mechanisms, models and physical applications},\
  }\href {https://doi.org/10.1016/0370-1573(90)90099-N} {\bibfield  {journal}
  {\bibinfo  {journal} {Physics reports}\ }\textbf {\bibinfo {volume} {195}},\
  \bibinfo {pages} {127} (\bibinfo {year} {1990})}\BibitemShut {NoStop}%
\bibitem [{\citenamefont {Cecconi}\ \emph {et~al.}(2022)\citenamefont
  {Cecconi}, \citenamefont {Costantini}, \citenamefont {Taloni},\ and\
  \citenamefont {Vulpiani}}]{cecconi2022probability}%
  \BibitemOpen
  \bibfield  {author} {\bibinfo {author} {\bibfnamefont {F.}~\bibnamefont
  {Cecconi}}, \bibinfo {author} {\bibfnamefont {G.}~\bibnamefont {Costantini}},
  \bibinfo {author} {\bibfnamefont {A.}~\bibnamefont {Taloni}},\ and\ \bibinfo
  {author} {\bibfnamefont {A.}~\bibnamefont {Vulpiani}},\ }\bibfield  {title}
  {\bibinfo {title} {Probability distribution functions of sub- and
  superdiffusive systems},\ }\href
  {https://doi.org/10.1103/PhysRevResearch.4.023192} {\bibfield  {journal}
  {\bibinfo  {journal} {Phys. Rev. Research}\ }\textbf {\bibinfo {volume}
  {4}},\ \bibinfo {pages} {023192} (\bibinfo {year} {2022})}\BibitemShut
  {NoStop}%
\bibitem [{\citenamefont {Klages}\ \emph {et~al.}(2008)\citenamefont {Klages},
  \citenamefont {Radons},\ and\ \citenamefont {Sokolov}}]{klages2008anomalous}%
  \BibitemOpen
  \bibfield  {author} {\bibinfo {author} {\bibfnamefont {R.}~\bibnamefont
  {Klages}}, \bibinfo {author} {\bibfnamefont {G.}~\bibnamefont {Radons}},\
  and\ \bibinfo {author} {\bibfnamefont {I.~M.}\ \bibnamefont {Sokolov}},\
  }\href@noop {} {\emph {\bibinfo {title} {Anomalous transport}}}\ (\bibinfo
  {publisher} {Wiley Online Library},\ \bibinfo {year} {2008})\BibitemShut
  {NoStop}%
\bibitem [{\citenamefont {Metzler}\ and\ \citenamefont
  {Klafter}(2000)}]{metzler2000random}%
  \BibitemOpen
  \bibfield  {author} {\bibinfo {author} {\bibfnamefont {R.}~\bibnamefont
  {Metzler}}\ and\ \bibinfo {author} {\bibfnamefont {J.}~\bibnamefont
  {Klafter}},\ }\bibfield  {title} {\bibinfo {title} {The random walk's guide
  to anomalous diffusion: a fractional dynamics approach},\ }\href
  {https://doi.org/https://doi.org/10.1016/S0370-1573(00)00070-3} {\bibfield
  {journal} {\bibinfo  {journal} {Physics Reports}\ }\textbf {\bibinfo {volume}
  {339}},\ \bibinfo {pages} {1} (\bibinfo {year} {2000})}\BibitemShut {NoStop}%
\bibitem [{\citenamefont {He}\ \emph {et~al.}(2008)\citenamefont {He},
  \citenamefont {Burov}, \citenamefont {Metzler},\ and\ \citenamefont
  {Barkai}}]{he2008random}%
  \BibitemOpen
  \bibfield  {author} {\bibinfo {author} {\bibfnamefont {Y.}~\bibnamefont
  {He}}, \bibinfo {author} {\bibfnamefont {S.}~\bibnamefont {Burov}}, \bibinfo
  {author} {\bibfnamefont {R.}~\bibnamefont {Metzler}},\ and\ \bibinfo {author}
  {\bibfnamefont {E.}~\bibnamefont {Barkai}},\ }\bibfield  {title} {\bibinfo
  {title} {Random time-scale invariant diffusion and transport coefficients},\
  }\href {https://doi.org/10.1103/PhysRevLett.101.058101} {\bibfield  {journal}
  {\bibinfo  {journal} {Phys. Rev. Lett.}\ }\textbf {\bibinfo {volume} {101}},\
  \bibinfo {pages} {058101} (\bibinfo {year} {2008})}\BibitemShut {NoStop}%
\bibitem [{\citenamefont {Sokolov}(2012)}]{sokolov2012models}%
  \BibitemOpen
  \bibfield  {author} {\bibinfo {author} {\bibfnamefont {I.~M.}\ \bibnamefont
  {Sokolov}},\ }\bibfield  {title} {\bibinfo {title} {Models of anomalous
  diffusion in crowded environments},\ }\href
  {https://doi.org/10.1039/C2SM25701G} {\bibfield  {journal} {\bibinfo
  {journal} {Soft Matter}\ }\textbf {\bibinfo {volume} {8}},\ \bibinfo {pages}
  {9043} (\bibinfo {year} {2012})}\BibitemShut {NoStop}%
\bibitem [{\citenamefont {Golding}\ and\ \citenamefont
  {Cox}(2006)}]{golding2006physical}%
  \BibitemOpen
  \bibfield  {author} {\bibinfo {author} {\bibfnamefont {I.}~\bibnamefont
  {Golding}}\ and\ \bibinfo {author} {\bibfnamefont {E.~C.}\ \bibnamefont
  {Cox}},\ }\bibfield  {title} {\bibinfo {title} {Physical nature of bacterial
  cytoplasm},\ }\href {https://doi.org/10.1103/PhysRevLett.96.098102}
  {\bibfield  {journal} {\bibinfo  {journal} {Phys. Rev. Lett.}\ }\textbf
  {\bibinfo {volume} {96}},\ \bibinfo {pages} {098102} (\bibinfo {year}
  {2006})}\BibitemShut {NoStop}%
\bibitem [{\citenamefont {Lubelski}\ \emph {et~al.}(2008)\citenamefont
  {Lubelski}, \citenamefont {Sokolov},\ and\ \citenamefont
  {Klafter}}]{lubelski2008nonergodicity}%
  \BibitemOpen
  \bibfield  {author} {\bibinfo {author} {\bibfnamefont {A.}~\bibnamefont
  {Lubelski}}, \bibinfo {author} {\bibfnamefont {I.~M.}\ \bibnamefont
  {Sokolov}},\ and\ \bibinfo {author} {\bibfnamefont {J.}~\bibnamefont
  {Klafter}},\ }\bibfield  {title} {\bibinfo {title} {Nonergodicity mimics
  inhomogeneity in single particle tracking},\ }\href
  {https://doi.org/10.1103/PhysRevLett.100.250602} {\bibfield  {journal}
  {\bibinfo  {journal} {Phys. Rev. Lett.}\ }\textbf {\bibinfo {volume} {100}},\
  \bibinfo {pages} {250602} (\bibinfo {year} {2008})}\BibitemShut {NoStop}%
\bibitem [{\citenamefont {Weber}\ \emph {et~al.}(2010)\citenamefont {Weber},
  \citenamefont {Spakowitz},\ and\ \citenamefont
  {Theriot}}]{weber2010bacterial}%
  \BibitemOpen
  \bibfield  {author} {\bibinfo {author} {\bibfnamefont {S.~C.}\ \bibnamefont
  {Weber}}, \bibinfo {author} {\bibfnamefont {A.~J.}\ \bibnamefont
  {Spakowitz}},\ and\ \bibinfo {author} {\bibfnamefont {J.~A.}\ \bibnamefont
  {Theriot}},\ }\bibfield  {title} {\bibinfo {title} {Bacterial chromosomal
  loci move subdiffusively through a viscoelastic cytoplasm},\ }\href
  {https://doi.org/10.1103/PhysRevLett.104.238102} {\bibfield  {journal}
  {\bibinfo  {journal} {Phys. Rev. Lett.}\ }\textbf {\bibinfo {volume} {104}},\
  \bibinfo {pages} {238102} (\bibinfo {year} {2010})}\BibitemShut {NoStop}%
\bibitem [{\citenamefont {Weigel}\ \emph {et~al.}(2011)\citenamefont {Weigel},
  \citenamefont {Simon}, \citenamefont {Tamkun},\ and\ \citenamefont
  {Krapf}}]{weigel2011ergodic}%
  \BibitemOpen
  \bibfield  {author} {\bibinfo {author} {\bibfnamefont {A.~V.}\ \bibnamefont
  {Weigel}}, \bibinfo {author} {\bibfnamefont {B.}~\bibnamefont {Simon}},
  \bibinfo {author} {\bibfnamefont {M.~M.}\ \bibnamefont {Tamkun}},\ and\
  \bibinfo {author} {\bibfnamefont {D.}~\bibnamefont {Krapf}},\ }\bibfield
  {title} {\bibinfo {title} {Ergodic and nonergodic processes coexist in the
  plasma membrane as observed by single-molecule tracking},\ }\href
  {https://doi.org/10.1073/pnas.1016325108} {\bibfield  {journal} {\bibinfo
  {journal} {Proceedings of the National Academy of Sciences}\ }\textbf
  {\bibinfo {volume} {108}},\ \bibinfo {pages} {6438} (\bibinfo {year}
  {2011})}\BibitemShut {NoStop}%
\bibitem [{\citenamefont {Weber}\ \emph {et~al.}(2012)\citenamefont {Weber},
  \citenamefont {Spakowitz},\ and\ \citenamefont
  {Theriot}}]{weber2012nonthermal}%
  \BibitemOpen
  \bibfield  {author} {\bibinfo {author} {\bibfnamefont {S.~C.}\ \bibnamefont
  {Weber}}, \bibinfo {author} {\bibfnamefont {A.~J.}\ \bibnamefont
  {Spakowitz}},\ and\ \bibinfo {author} {\bibfnamefont {J.~A.}\ \bibnamefont
  {Theriot}},\ }\bibfield  {title} {\bibinfo {title} {Nonthermal atp-dependent
  fluctuations contribute to the in vivo motion of chromosomal loci},\ }\href
  {https://doi.org/10.1073/pnas.1119505109} {\bibfield  {journal} {\bibinfo
  {journal} {Proceedings of the National Academy of Sciences}\ }\textbf
  {\bibinfo {volume} {109}},\ \bibinfo {pages} {7338} (\bibinfo {year}
  {2012})}\BibitemShut {NoStop}%
\bibitem [{\citenamefont {Viswanathan}\ \emph {et~al.}(1999)\citenamefont
  {Viswanathan}, \citenamefont {Buldyrev}, \citenamefont {Havlin},
  \citenamefont {Da~Luz}, \citenamefont {Raposo},\ and\ \citenamefont
  {Stanley}}]{viswanathan1999optimizing}%
  \BibitemOpen
  \bibfield  {author} {\bibinfo {author} {\bibfnamefont {G.~M.}\ \bibnamefont
  {Viswanathan}}, \bibinfo {author} {\bibfnamefont {S.~V.}\ \bibnamefont
  {Buldyrev}}, \bibinfo {author} {\bibfnamefont {S.}~\bibnamefont {Havlin}},
  \bibinfo {author} {\bibfnamefont {M.}~\bibnamefont {Da~Luz}}, \bibinfo
  {author} {\bibfnamefont {E.}~\bibnamefont {Raposo}},\ and\ \bibinfo {author}
  {\bibfnamefont {H.~E.}\ \bibnamefont {Stanley}},\ }\bibfield  {title}
  {\bibinfo {title} {Optimizing the success of random searches},\ }\href
  {https://doi.org/10.1038/44831} {\bibfield  {journal} {\bibinfo  {journal}
  {nature}\ }\textbf {\bibinfo {volume} {401}},\ \bibinfo {pages} {911}
  (\bibinfo {year} {1999})}\BibitemShut {NoStop}%
\bibitem [{\citenamefont {Barthelemy}\ \emph {et~al.}(2008)\citenamefont
  {Barthelemy}, \citenamefont {Bertolotti},\ and\ \citenamefont
  {Wiersma}}]{barthelemy2008levy}%
  \BibitemOpen
  \bibfield  {author} {\bibinfo {author} {\bibfnamefont {P.}~\bibnamefont
  {Barthelemy}}, \bibinfo {author} {\bibfnamefont {J.}~\bibnamefont
  {Bertolotti}},\ and\ \bibinfo {author} {\bibfnamefont {D.~S.}\ \bibnamefont
  {Wiersma}},\ }\bibfield  {title} {\bibinfo {title} {A l{\'e}vy flight for
  light},\ }\href {https://doi.org/10.1038/nature06948} {\bibfield  {journal}
  {\bibinfo  {journal} {Nature}\ }\textbf {\bibinfo {volume} {453}},\ \bibinfo
  {pages} {495} (\bibinfo {year} {2008})}\BibitemShut {NoStop}%
\bibitem [{\citenamefont {Lagutin}\ and\ \citenamefont
  {Uchaikin}(2003)}]{lagutin2003anomalous}%
  \BibitemOpen
  \bibfield  {author} {\bibinfo {author} {\bibfnamefont {A.}~\bibnamefont
  {Lagutin}}\ and\ \bibinfo {author} {\bibfnamefont {V.}~\bibnamefont
  {Uchaikin}},\ }\bibfield  {title} {\bibinfo {title} {Anomalous diffusion
  equation: Application to cosmic ray transport},\ }\href
  {https://doi.org/10.1016/S0168-583X(02)01362-9} {\bibfield  {journal}
  {\bibinfo  {journal} {Nuclear Instruments and Methods in Physics Research
  Section B: Beam Interactions with Materials and Atoms}\ }\textbf {\bibinfo
  {volume} {201}},\ \bibinfo {pages} {212} (\bibinfo {year}
  {2003})}\BibitemShut {NoStop}%
\bibitem [{\citenamefont {Uchaikin}(2013)}]{uchaikin2013fractional}%
  \BibitemOpen
  \bibfield  {author} {\bibinfo {author} {\bibfnamefont {V.~V.}\ \bibnamefont
  {Uchaikin}},\ }\bibfield  {title} {\bibinfo {title} {Fractional phenomenology
  of cosmic ray anomalous diffusion},\ }\href
  {https://doi.org/10.3367/ufne.0183.201311b.1175} {\bibfield  {journal}
  {\bibinfo  {journal} {Physics-Uspekhi}\ }\textbf {\bibinfo {volume} {56}},\
  \bibinfo {pages} {1074} (\bibinfo {year} {2013})}\BibitemShut {NoStop}%
\bibitem [{\citenamefont {Havlin}\ and\ \citenamefont
  {Ben-Avraham}(1987)}]{havlin1987diffusion}%
  \BibitemOpen
  \bibfield  {author} {\bibinfo {author} {\bibfnamefont {S.}~\bibnamefont
  {Havlin}}\ and\ \bibinfo {author} {\bibfnamefont {D.}~\bibnamefont
  {Ben-Avraham}},\ }\bibfield  {title} {\bibinfo {title} {Diffusion in
  disordered media},\ }\href {https://doi.org/10.1080/00018738700101072}
  {\bibfield  {journal} {\bibinfo  {journal} {Advances in Physics}\ }\textbf
  {\bibinfo {volume} {36}},\ \bibinfo {pages} {695} (\bibinfo {year} {1987})},\
  \Eprint {https://arxiv.org/abs/https://doi.org/10.1080/00018738700101072}
  {https://doi.org/10.1080/00018738700101072} \BibitemShut {NoStop}%
\bibitem [{\citenamefont {Havlin}\ \emph {et~al.}(1985)\citenamefont {Havlin},
  \citenamefont {Movshovitz}, \citenamefont {Trus},\ and\ \citenamefont
  {Weiss}}]{Havlin1985}%
  \BibitemOpen
  \bibfield  {author} {\bibinfo {author} {\bibfnamefont {S.}~\bibnamefont
  {Havlin}}, \bibinfo {author} {\bibfnamefont {D.}~\bibnamefont {Movshovitz}},
  \bibinfo {author} {\bibfnamefont {B.}~\bibnamefont {Trus}},\ and\ \bibinfo
  {author} {\bibfnamefont {G.~H.}\ \bibnamefont {Weiss}},\ }\bibfield  {title}
  {\bibinfo {title} {Probability densities for the displacement of random walks
  on percolation clusters},\ }\href
  {https://doi.org/10.1088/0305-4470/18/12/006} {\bibfield  {journal} {\bibinfo
   {journal} {Journal of Physics A: Mathematical and General}\ }\textbf
  {\bibinfo {volume} {18}},\ \bibinfo {pages} {L719} (\bibinfo {year}
  {1985})}\BibitemShut {NoStop}%
\bibitem [{\citenamefont {Guyer}(1985)}]{Guyer1985}%
  \BibitemOpen
  \bibfield  {author} {\bibinfo {author} {\bibfnamefont {R.~A.}\ \bibnamefont
  {Guyer}},\ }\bibfield  {title} {\bibinfo {title} {Diffusive motion on a
  fractal; ${G}_{\mathrm{nm}}$(t)},\ }\href
  {https://doi.org/10.1103/PhysRevA.32.2324} {\bibfield  {journal} {\bibinfo
  {journal} {Phys. Rev. A}\ }\textbf {\bibinfo {volume} {32}},\ \bibinfo
  {pages} {2324} (\bibinfo {year} {1985})}\BibitemShut {NoStop}%
\bibitem [{\citenamefont {McCoy}\ and\ \citenamefont
  {Wu}(2014)}]{mccoy2014two}%
  \BibitemOpen
  \bibfield  {author} {\bibinfo {author} {\bibfnamefont {B.~M.}\ \bibnamefont
  {McCoy}}\ and\ \bibinfo {author} {\bibfnamefont {T.~T.}\ \bibnamefont {Wu}},\
  }\href@noop {} {\emph {\bibinfo {title} {The two-dimensional Ising model}}}\
  (\bibinfo  {publisher} {Courier Corporation},\ \bibinfo {year}
  {2014})\BibitemShut {NoStop}%
\bibitem [{\citenamefont {Eisenriegler}\ and\ \citenamefont
  {Tomaschitz}(1987)}]{eisenriegler1987}%
  \BibitemOpen
  \bibfield  {author} {\bibinfo {author} {\bibfnamefont {E.}~\bibnamefont
  {Eisenriegler}}\ and\ \bibinfo {author} {\bibfnamefont {R.}~\bibnamefont
  {Tomaschitz}},\ }\bibfield  {title} {\bibinfo {title} {Helmholtz free energy
  of finite spin systems near criticality},\ }\href@noop {} {\bibfield
  {journal} {\bibinfo  {journal} {Physical Review B}\ }\textbf {\bibinfo
  {volume} {35}},\ \bibinfo {pages} {4876} (\bibinfo {year}
  {1987})}\BibitemShut {NoStop}%
\bibitem [{\citenamefont {Hilfer}(1994)}]{hilfer1994}%
  \BibitemOpen
  \bibfield  {author} {\bibinfo {author} {\bibfnamefont {R.}~\bibnamefont
  {Hilfer}},\ }\bibfield  {title} {\bibinfo {title} {Absence of hyperscaling
  violations for phase transitions with positive specific heat exponent},\
  }\href@noop {} {\bibfield  {journal} {\bibinfo  {journal} {Zeitschrift
  f{\"u}r Physik B Condensed Matter}\ }\textbf {\bibinfo {volume} {96}},\
  \bibinfo {pages} {63} (\bibinfo {year} {1994})}\BibitemShut {NoStop}%
\bibitem [{\citenamefont {Kadanoff}\ \emph {et~al.}(1967)\citenamefont
  {Kadanoff}, \citenamefont {G\"otze}, \citenamefont {Hamblen}, \citenamefont
  {Hecht}, \citenamefont {Lewis}, \citenamefont {Pakciauskas}, \citenamefont
  {Rayl}, \citenamefont {Swift}, \citenamefont {Aspnes},\ and\ \citenamefont
  {Kane}}]{kadanoff1967static}%
  \BibitemOpen
  \bibfield  {author} {\bibinfo {author} {\bibfnamefont {L.~P.}\ \bibnamefont
  {Kadanoff}}, \bibinfo {author} {\bibfnamefont {W.}~\bibnamefont {G\"otze}},
  \bibinfo {author} {\bibfnamefont {D.}~\bibnamefont {Hamblen}}, \bibinfo
  {author} {\bibfnamefont {R.}~\bibnamefont {Hecht}}, \bibinfo {author}
  {\bibfnamefont {E.~A.~S.}\ \bibnamefont {Lewis}}, \bibinfo {author}
  {\bibfnamefont {V.~V.}\ \bibnamefont {Pakciauskas}}, \bibinfo {author}
  {\bibfnamefont {M.}~\bibnamefont {Rayl}}, \bibinfo {author} {\bibfnamefont
  {J.}~\bibnamefont {Swift}}, \bibinfo {author} {\bibfnamefont
  {D.}~\bibnamefont {Aspnes}},\ and\ \bibinfo {author} {\bibfnamefont
  {J.}~\bibnamefont {Kane}},\ }\bibfield  {title} {\bibinfo {title} {Static
  phenomena near critical points: Theory and experiment},\ }\href
  {https://doi.org/10.1103/RevModPhys.39.395} {\bibfield  {journal} {\bibinfo
  {journal} {Rev. Mod. Phys.}\ }\textbf {\bibinfo {volume} {39}},\ \bibinfo
  {pages} {395} (\bibinfo {year} {1967})}\BibitemShut {NoStop}%
\bibitem [{\citenamefont {Bruce}(1995)}]{Bruce1995}%
  \BibitemOpen
  \bibfield  {author} {\bibinfo {author} {\bibfnamefont {A.}~\bibnamefont
  {Bruce}},\ }\bibfield  {title} {\bibinfo {title} {Critical finite-size
  scaling of the free energy},\ }\href
  {https://doi.org/10.1088/0305-4470/28/12/008} {\bibfield  {journal} {\bibinfo
   {journal} {Journal of Physics A: Mathematical and General}\ }\textbf
  {\bibinfo {volume} {28}},\ \bibinfo {pages} {3345} (\bibinfo {year}
  {1995})}\BibitemShut {NoStop}%
\bibitem [{\citenamefont {Hilfer}\ and\ \citenamefont
  {Wilding}(1995)}]{Hilfer1995}%
  \BibitemOpen
  \bibfield  {author} {\bibinfo {author} {\bibfnamefont {R.}~\bibnamefont
  {Hilfer}}\ and\ \bibinfo {author} {\bibfnamefont {N.}~\bibnamefont
  {Wilding}},\ }\bibfield  {title} {\bibinfo {title} {Are critical finite-size
  scaling functions calculable from knowledge of an appropriate critical
  exponent?},\ }\href {https://doi.org/10.1088/0305-4470/28/10/001} {\bibfield
  {journal} {\bibinfo  {journal} {Journal of Physics A: Mathematical and
  General}\ }\textbf {\bibinfo {volume} {28}},\ \bibinfo {pages} {L281}
  (\bibinfo {year} {1995})}\BibitemShut {NoStop}%
\bibitem [{\citenamefont {Stauffer}(1998)}]{stauffer1998}%
  \BibitemOpen
  \bibfield  {author} {\bibinfo {author} {\bibfnamefont {D.}~\bibnamefont
  {Stauffer}},\ }\bibfield  {title} {\bibinfo {title} {Monte carlo
  investigation of rare magnetization fluctuations in ising models},\ }\href
  {https://doi.org/10.1142/S0129183198000510} {\bibfield  {journal} {\bibinfo
  {journal} {International Journal of Modern Physics C}\ }\textbf {\bibinfo
  {volume} {9}},\ \bibinfo {pages} {625} (\bibinfo {year} {1998})}\BibitemShut
  {NoStop}%
\bibitem [{\citenamefont {Tsypin}\ and\ \citenamefont
  {Bl\"ote}(2000)}]{tsypin2000probability}%
  \BibitemOpen
  \bibfield  {author} {\bibinfo {author} {\bibfnamefont {M.~M.}\ \bibnamefont
  {Tsypin}}\ and\ \bibinfo {author} {\bibfnamefont {H.~W.~J.}\ \bibnamefont
  {Bl\"ote}},\ }\bibfield  {title} {\bibinfo {title} {Probability distribution
  of the order parameter for the three-dimensional ising-model universality
  class: A high-precision monte carlo study},\ }\href
  {https://doi.org/10.1103/PhysRevE.62.73} {\bibfield  {journal} {\bibinfo
  {journal} {Phys. Rev. E}\ }\textbf {\bibinfo {volume} {62}},\ \bibinfo
  {pages} {73} (\bibinfo {year} {2000})}\BibitemShut {NoStop}%
\bibitem [{\citenamefont {Berg}\ \emph {et~al.}(2002)\citenamefont {Berg},
  \citenamefont {Billoire},\ and\ \citenamefont {Janke}}]{berg2002overlap}%
  \BibitemOpen
  \bibfield  {author} {\bibinfo {author} {\bibfnamefont {B.~A.}\ \bibnamefont
  {Berg}}, \bibinfo {author} {\bibfnamefont {A.}~\bibnamefont {Billoire}},\
  and\ \bibinfo {author} {\bibfnamefont {W.}~\bibnamefont {Janke}},\ }\bibfield
   {title} {\bibinfo {title} {Overlap distribution of the three-dimensional
  ising model},\ }\href {https://doi.org/10.1103/PhysRevE.66.046122} {\bibfield
   {journal} {\bibinfo  {journal} {Phys. Rev. E}\ }\textbf {\bibinfo {volume}
  {66}},\ \bibinfo {pages} {046122} (\bibinfo {year} {2002})}\BibitemShut
  {NoStop}%
\bibitem [{\citenamefont {Hilfer}\ \emph {et~al.}(2003)\citenamefont {Hilfer},
  \citenamefont {Biswal}, \citenamefont {Mattutis},\ and\ \citenamefont
  {Janke}}]{hilfer2003multicanonical}%
  \BibitemOpen
  \bibfield  {author} {\bibinfo {author} {\bibfnamefont {R.}~\bibnamefont
  {Hilfer}}, \bibinfo {author} {\bibfnamefont {B.}~\bibnamefont {Biswal}},
  \bibinfo {author} {\bibfnamefont {H.~G.}\ \bibnamefont {Mattutis}},\ and\
  \bibinfo {author} {\bibfnamefont {W.}~\bibnamefont {Janke}},\ }\bibfield
  {title} {\bibinfo {title} {Multicanonical monte carlo study and analysis of
  tails for the order-parameter distribution of the two-dimensional ising
  model},\ }\href {https://doi.org/10.1103/PhysRevE.68.046123} {\bibfield
  {journal} {\bibinfo  {journal} {Phys. Rev. E}\ }\textbf {\bibinfo {volume}
  {68}},\ \bibinfo {pages} {046123} (\bibinfo {year} {2003})}\BibitemShut
  {NoStop}%
\bibitem [{\citenamefont {Hilfer}\ \emph {et~al.}(2005)\citenamefont {Hilfer},
  \citenamefont {Biswal}, \citenamefont {Mattutis},\ and\ \citenamefont
  {Janke}}]{hilfer2005multicanonical}%
  \BibitemOpen
  \bibfield  {author} {\bibinfo {author} {\bibfnamefont {R.}~\bibnamefont
  {Hilfer}}, \bibinfo {author} {\bibfnamefont {B.}~\bibnamefont {Biswal}},
  \bibinfo {author} {\bibfnamefont {H.-G.}\ \bibnamefont {Mattutis}},\ and\
  \bibinfo {author} {\bibfnamefont {W.}~\bibnamefont {Janke}},\ }\bibfield
  {title} {\bibinfo {title} {Multicanonical simulations of the tails of the
  order-parameter distribution of the two-dimensional ising model},\ }\href
  {https://doi.org/10.1016/j.cpc.2005.03.053} {\bibfield  {journal} {\bibinfo
  {journal} {Computer physics communications}\ }\textbf {\bibinfo {volume}
  {169}},\ \bibinfo {pages} {230} (\bibinfo {year} {2005})}\BibitemShut
  {NoStop}%
\bibitem [{\citenamefont {Touchette}(2009)}]{Touchette2009}%
  \BibitemOpen
  \bibfield  {author} {\bibinfo {author} {\bibfnamefont {H.}~\bibnamefont
  {Touchette}},\ }\bibfield  {title} {\bibinfo {title} {The large deviation
  approach to statistical mechanics},\ }\href
  {https://doi.org/https://doi.org/10.1016/j.physrep.2009.05.002} {\bibfield
  {journal} {\bibinfo  {journal} {Physics Reports}\ }\textbf {\bibinfo {volume}
  {478}},\ \bibinfo {pages} {1} (\bibinfo {year} {2009})}\BibitemShut {NoStop}%
\bibitem [{\citenamefont {Touchette}\ and\ \citenamefont
  {Harris}(2013)}]{touchette2013}%
  \BibitemOpen
  \bibfield  {author} {\bibinfo {author} {\bibfnamefont {H.}~\bibnamefont
  {Touchette}}\ and\ \bibinfo {author} {\bibfnamefont {R.~J.}\ \bibnamefont
  {Harris}},\ }\bibinfo {title} {Large deviation approach to nonequilibrium
  systems},\ in\ \href {https://doi.org/10.1002/9783527658701.ch11} {\emph
  {\bibinfo {booktitle} {Nonequilibrium Statistical Physics of Small
  Systems}}},\ \bibinfo {editor} {edited by\ \bibinfo {editor} {\bibfnamefont
  {R.}~\bibnamefont {Klages}}, \bibinfo {editor} {\bibfnamefont
  {W.}~\bibnamefont {Just}},\ and\ \bibinfo {editor} {\bibfnamefont
  {C.}~\bibnamefont {Jarzynski}}}\ (\bibinfo  {publisher} {John Wiley \& Sons,
  Ltd},\ \bibinfo {year} {2013})\ Chap.~\bibinfo {chapter} {11}, pp.\ \bibinfo
  {pages} {335--360}\BibitemShut {NoStop}%
\bibitem [{\citenamefont {Teza}\ \emph {et~al.}(2022)\citenamefont {Teza},
  \citenamefont {Yaacoby},\ and\ \citenamefont {Raz}}]{teza2022eigenvalue}%
  \BibitemOpen
  \bibfield  {author} {\bibinfo {author} {\bibfnamefont {G.}~\bibnamefont
  {Teza}}, \bibinfo {author} {\bibfnamefont {R.}~\bibnamefont {Yaacoby}},\ and\
  \bibinfo {author} {\bibfnamefont {O.}~\bibnamefont {Raz}},\ }\bibfield
  {title} {\bibinfo {title} {Eigenvalue crossing as a phase transition in
  relaxation dynamics},\ }\bibfield  {journal} {\bibinfo  {journal} {arXiv
  preprint arXiv:2209.09307}\ }\href
  {https://doi.org/10.48550/arXiv.2209.09307} {10.48550/arXiv.2209.09307}
  (\bibinfo {year} {2022})\BibitemShut {NoStop}%
\bibitem [{\citenamefont {Bender}\ and\ \citenamefont
  {Orszag}(1978)}]{bender1978}%
  \BibitemOpen
  \bibfield  {author} {\bibinfo {author} {\bibfnamefont {C.~M.}\ \bibnamefont
  {Bender}}\ and\ \bibinfo {author} {\bibfnamefont {S.~A.}\ \bibnamefont
  {Orszag}},\ }\bibfield  {title} {\bibinfo {title} {Advanced mathematical
  methods for scientists and engineers},\ }\href
  {https://doi.org/10.1007/978-1-4757-3069-2} {\bibfield  {journal} {\bibinfo
  {journal} {McGraw-Hill, New York}\ }\textbf {\bibinfo {volume} {1}},\
  \bibinfo {pages} {14} (\bibinfo {year} {1978})}\BibitemShut {NoStop}%
\bibitem [{SM()}]{SM}%
  \BibitemOpen
  \href@noop {} {\bibinfo {title} {See supplemental material at ... for
  additional details of the calculations at the basis of the results presented
  in the main text.}}\BibitemShut {Stop}%
\bibitem [{\citenamefont {Montroll}\ and\ \citenamefont
  {Weiss}(1965)}]{montroll1965random}%
  \BibitemOpen
  \bibfield  {author} {\bibinfo {author} {\bibfnamefont {E.~W.}\ \bibnamefont
  {Montroll}}\ and\ \bibinfo {author} {\bibfnamefont {G.~H.}\ \bibnamefont
  {Weiss}},\ }\bibfield  {title} {\bibinfo {title} {Random walks on lattices.
  ii},\ }\href {https://doi.org/10.1063/1.1704269} {\bibfield  {journal}
  {\bibinfo  {journal} {Journal of Mathematical Physics}\ }\textbf {\bibinfo
  {volume} {6}},\ \bibinfo {pages} {167} (\bibinfo {year} {1965})}\BibitemShut
  {NoStop}%
\bibitem [{\citenamefont {Kenkre}\ \emph {et~al.}(1973)\citenamefont {Kenkre},
  \citenamefont {Montroll},\ and\ \citenamefont
  {Shlesinger}}]{kenkre1973generalized}%
  \BibitemOpen
  \bibfield  {author} {\bibinfo {author} {\bibfnamefont {V.}~\bibnamefont
  {Kenkre}}, \bibinfo {author} {\bibfnamefont {E.}~\bibnamefont {Montroll}},\
  and\ \bibinfo {author} {\bibfnamefont {M.}~\bibnamefont {Shlesinger}},\
  }\bibfield  {title} {\bibinfo {title} {Generalized master equations for
  continuous-time random walks},\ }\href {https://doi.org/10.1007/BF01016796}
  {\bibfield  {journal} {\bibinfo  {journal} {Journal of Statistical Physics}\
  }\textbf {\bibinfo {volume} {9}},\ \bibinfo {pages} {45} (\bibinfo {year}
  {1973})}\BibitemShut {NoStop}%
\bibitem [{\citenamefont {Carpinteri}\ and\ \citenamefont
  {Mainardi}(2014)}]{carpinteri2014fractals}%
  \BibitemOpen
  \bibfield  {author} {\bibinfo {author} {\bibfnamefont {A.}~\bibnamefont
  {Carpinteri}}\ and\ \bibinfo {author} {\bibfnamefont {F.}~\bibnamefont
  {Mainardi}},\ }\href@noop {} {\emph {\bibinfo {title} {Fractals and
  fractional calculus in continuum mechanics}}},\ Vol.\ \bibinfo {volume}
  {378}\ (\bibinfo  {publisher} {Springer},\ \bibinfo {year}
  {2014})\BibitemShut {NoStop}%
\bibitem [{\citenamefont {Teza}\ and\ \citenamefont
  {Stella}(2020)}]{teza2020exact}%
  \BibitemOpen
  \bibfield  {author} {\bibinfo {author} {\bibfnamefont {G.}~\bibnamefont
  {Teza}}\ and\ \bibinfo {author} {\bibfnamefont {A.~L.}\ \bibnamefont
  {Stella}},\ }\bibfield  {title} {\bibinfo {title} {Exact coarse graining
  preserves entropy production out of equilibrium},\ }\href
  {https://doi.org/10.1103/PhysRevLett.125.110601} {\bibfield  {journal}
  {\bibinfo  {journal} {Phys. Rev. Lett.}\ }\textbf {\bibinfo {volume} {125}},\
  \bibinfo {pages} {110601} (\bibinfo {year} {2020})}\BibitemShut {NoStop}%
\bibitem [{\citenamefont {Teza}(2020)}]{teza2020thesis}%
  \BibitemOpen
  \bibfield  {author} {\bibinfo {author} {\bibfnamefont {G.}~\bibnamefont
  {Teza}},\ }\emph {\bibinfo {title} {Out of equilibrium dynamics: from an
  entropy of the growth to the growth of entropy production}},\ \href
  {http://paduaresearch.cab.unipd.it/12995/} {Ph.D. thesis},\ \bibinfo
  {school} {University of Padova} (\bibinfo {year} {2020})\BibitemShut
  {NoStop}%
\bibitem [{\citenamefont {Gorenflo}\ \emph {et~al.}(2020)\citenamefont
  {Gorenflo}, \citenamefont {Kilbas}, \citenamefont {Mainardi},\ and\
  \citenamefont {Rogosin}}]{gorenflo2020mittag}%
  \BibitemOpen
  \bibfield  {author} {\bibinfo {author} {\bibfnamefont {R.}~\bibnamefont
  {Gorenflo}}, \bibinfo {author} {\bibfnamefont {A.~A.}\ \bibnamefont
  {Kilbas}}, \bibinfo {author} {\bibfnamefont {F.}~\bibnamefont {Mainardi}},\
  and\ \bibinfo {author} {\bibfnamefont {S.~V.}\ \bibnamefont {Rogosin}},\
  }\href@noop {} {\emph {\bibinfo {title} {Mittag-Leffler functions, related
  topics and applications}}}\ (\bibinfo  {publisher} {Springer},\ \bibinfo
  {year} {2020})\BibitemShut {NoStop}%
\bibitem [{\citenamefont {Schneider}\ and\ \citenamefont
  {Wyss}(1989)}]{schneider1989fractional}%
  \BibitemOpen
  \bibfield  {author} {\bibinfo {author} {\bibfnamefont {W.~R.}\ \bibnamefont
  {Schneider}}\ and\ \bibinfo {author} {\bibfnamefont {W.}~\bibnamefont
  {Wyss}},\ }\bibfield  {title} {\bibinfo {title} {Fractional diffusion and
  wave equations},\ }\href {https://doi.org/10.1063/1.528578} {\bibfield
  {journal} {\bibinfo  {journal} {Journal of Mathematical Physics}\ }\textbf
  {\bibinfo {volume} {30}},\ \bibinfo {pages} {134} (\bibinfo {year}
  {1989})}\BibitemShut {NoStop}%
\bibitem [{\citenamefont {Mainardi}\ and\ \citenamefont
  {Tomirotti}(1994)}]{mainardi1994special}%
  \BibitemOpen
  \bibfield  {author} {\bibinfo {author} {\bibfnamefont {F.}~\bibnamefont
  {Mainardi}}\ and\ \bibinfo {author} {\bibfnamefont {M.}~\bibnamefont
  {Tomirotti}},\ }\bibfield  {title} {\bibinfo {title} {On a special function
  arising in the time fractional diffusion-wave equation},\ }\href@noop {}
  {\bibfield  {journal} {\bibinfo  {journal} {Transform Methods and Special
  Functions, Sofia}\ }\textbf {\bibinfo {volume} {171}} (\bibinfo {year}
  {1994})}\BibitemShut {NoStop}%
\bibitem [{\citenamefont {Barkai}\ \emph {et~al.}(2000)\citenamefont {Barkai},
  \citenamefont {Metzler},\ and\ \citenamefont
  {Klafter}}]{barkai2000continuous}%
  \BibitemOpen
  \bibfield  {author} {\bibinfo {author} {\bibfnamefont {E.}~\bibnamefont
  {Barkai}}, \bibinfo {author} {\bibfnamefont {R.}~\bibnamefont {Metzler}},\
  and\ \bibinfo {author} {\bibfnamefont {J.}~\bibnamefont {Klafter}},\
  }\bibfield  {title} {\bibinfo {title} {From continuous time random walks to
  the fractional fokker-planck equation},\ }\href
  {https://doi.org/10.1103/PhysRevE.61.132} {\bibfield  {journal} {\bibinfo
  {journal} {Phys. Rev. E}\ }\textbf {\bibinfo {volume} {61}},\ \bibinfo
  {pages} {132} (\bibinfo {year} {2000})}\BibitemShut {NoStop}%
\bibitem [{\citenamefont {Mainardi}\ \emph {et~al.}(2010)\citenamefont
  {Mainardi}, \citenamefont {Mura},\ and\ \citenamefont
  {Pagnini}}]{mainardi2010wright}%
  \BibitemOpen
  \bibfield  {author} {\bibinfo {author} {\bibfnamefont {F.}~\bibnamefont
  {Mainardi}}, \bibinfo {author} {\bibfnamefont {A.}~\bibnamefont {Mura}},\
  and\ \bibinfo {author} {\bibfnamefont {G.}~\bibnamefont {Pagnini}},\
  }\bibfield  {title} {\bibinfo {title} {The {M}-{W}right function in
  time-fractional diffusion processes: a tutorial survey},\ }\bibfield
  {journal} {\bibinfo  {journal} {International Journal of Differential
  Equations}\ }\textbf {\bibinfo {volume} {2010}},\ \href
  {https://doi.org/10.1155/2010/104505} {10.1155/2010/104505} (\bibinfo {year}
  {2010})\BibitemShut {NoStop}%
\bibitem [{\citenamefont {Katz}\ \emph {et~al.}(1983)\citenamefont {Katz},
  \citenamefont {Lebowitz},\ and\ \citenamefont {Spohn}}]{katz1983phase}%
  \BibitemOpen
  \bibfield  {author} {\bibinfo {author} {\bibfnamefont {S.}~\bibnamefont
  {Katz}}, \bibinfo {author} {\bibfnamefont {J.~L.}\ \bibnamefont {Lebowitz}},\
  and\ \bibinfo {author} {\bibfnamefont {H.}~\bibnamefont {Spohn}},\ }\bibfield
   {title} {\bibinfo {title} {Phase transitions in stationary nonequilibrium
  states of model lattice systems},\ }\href
  {https://doi.org/10.1103/PhysRevB.28.1655} {\bibfield  {journal} {\bibinfo
  {journal} {Phys. Rev. B}\ }\textbf {\bibinfo {volume} {28}},\ \bibinfo
  {pages} {1655} (\bibinfo {year} {1983})}\BibitemShut {NoStop}%
\bibitem [{\citenamefont {Teza}\ \emph {et~al.}(2020)\citenamefont {Teza},
  \citenamefont {Iubini}, \citenamefont {Baiesi}, \citenamefont {Stella},\ and\
  \citenamefont {Vanderzande}}]{teza2020rate}%
  \BibitemOpen
  \bibfield  {author} {\bibinfo {author} {\bibfnamefont {G.}~\bibnamefont
  {Teza}}, \bibinfo {author} {\bibfnamefont {S.}~\bibnamefont {Iubini}},
  \bibinfo {author} {\bibfnamefont {M.}~\bibnamefont {Baiesi}}, \bibinfo
  {author} {\bibfnamefont {A.~L.}\ \bibnamefont {Stella}},\ and\ \bibinfo
  {author} {\bibfnamefont {C.}~\bibnamefont {Vanderzande}},\ }\bibfield
  {title} {\bibinfo {title} {Rate dependence of current and fluctuations in
  jump models with negative differential mobility},\ }\href
  {https://doi.org/https://doi.org/10.1016/j.physa.2019.123176} {\bibfield
  {journal} {\bibinfo  {journal} {Physica A: Statistical Mechanics and its
  Applications}\ }\textbf {\bibinfo {volume} {552}},\ \bibinfo {pages} {123176}
  (\bibinfo {year} {2020})},\ \bibinfo {note} {tributes of Non-equilibrium
  Statistical Physics}\BibitemShut {NoStop}%
\bibitem [{\citenamefont {Chechkin}\ and\ \citenamefont
  {Klages}(2009)}]{chechkin2009fluctuation}%
  \BibitemOpen
  \bibfield  {author} {\bibinfo {author} {\bibfnamefont {A.~V.}\ \bibnamefont
  {Chechkin}}\ and\ \bibinfo {author} {\bibfnamefont {R.}~\bibnamefont
  {Klages}},\ }\bibfield  {title} {\bibinfo {title} {Fluctuation relations for
  anomalous dynamics},\ }\href
  {https://doi.org/10.1088/1742-5468/2009/03/l03002} {\bibfield  {journal}
  {\bibinfo  {journal} {Journal of Statistical Mechanics: Theory and
  Experiment}\ }\textbf {\bibinfo {volume} {2009}},\ \bibinfo {pages} {L03002}
  (\bibinfo {year} {2009})}\BibitemShut {NoStop}%
\bibitem [{\citenamefont {Sokolov}\ \emph {et~al.}(2009)\citenamefont
  {Sokolov}, \citenamefont {Heinsalu}, \citenamefont {Hänggi},\ and\
  \citenamefont {Goychuk}}]{sokolov2009universal}%
  \BibitemOpen
  \bibfield  {author} {\bibinfo {author} {\bibfnamefont {I.~M.}\ \bibnamefont
  {Sokolov}}, \bibinfo {author} {\bibfnamefont {E.}~\bibnamefont {Heinsalu}},
  \bibinfo {author} {\bibfnamefont {P.}~\bibnamefont {Hänggi}},\ and\ \bibinfo
  {author} {\bibfnamefont {I.}~\bibnamefont {Goychuk}},\ }\bibfield  {title}
  {\bibinfo {title} {Universal fluctuations in subdiffusive transport},\ }\href
  {https://doi.org/10.1209/0295-5075/86/30009} {\bibfield  {journal} {\bibinfo
  {journal} {{EPL} (Europhysics Letters)}\ }\textbf {\bibinfo {volume} {86}},\
  \bibinfo {pages} {30009} (\bibinfo {year} {2009})}\BibitemShut {NoStop}%
\bibitem [{\citenamefont {Dieterich}\ \emph {et~al.}(2015)\citenamefont
  {Dieterich}, \citenamefont {Klages},\ and\ \citenamefont
  {Chechkin}}]{dieterich2015fluctuation}%
  \BibitemOpen
  \bibfield  {author} {\bibinfo {author} {\bibfnamefont {P.}~\bibnamefont
  {Dieterich}}, \bibinfo {author} {\bibfnamefont {R.}~\bibnamefont {Klages}},\
  and\ \bibinfo {author} {\bibfnamefont {A.~V.}\ \bibnamefont {Chechkin}},\
  }\bibfield  {title} {\bibinfo {title} {Fluctuation relations for anomalous
  dynamics generated by time-fractional fokker{\textendash}planck equations},\
  }\href {https://doi.org/10.1088/1367-2630/17/7/075004} {\bibfield  {journal}
  {\bibinfo  {journal} {New Journal of Physics}\ }\textbf {\bibinfo {volume}
  {17}},\ \bibinfo {pages} {075004} (\bibinfo {year} {2015})}\BibitemShut
  {NoStop}%
\bibitem [{\citenamefont {Teza}\ and\ \citenamefont
  {Stella}()}]{teza2022renormalization}%
  \BibitemOpen
  \bibfield  {author} {\bibinfo {author} {\bibfnamefont {G.}~\bibnamefont
  {Teza}}\ and\ \bibinfo {author} {\bibfnamefont {A.~L.}\ \bibnamefont
  {Stella}},\ }\bibinfo {title} {(to be published)}\BibitemShut {NoStop}%
\bibitem [{\citenamefont {G\"{a}rtner}(1977)}]{gartner1977on}%
  \BibitemOpen
\bibfield  {title} {  }\bibfield  {author} {\bibinfo {author} {\bibfnamefont
  {J.}~\bibnamefont {G\"{a}rtner}},\ }\bibfield  {title} {\bibinfo {title} {On
  large deviations from the invariant measure},\ }\href
  {https://doi.org/10.1137/1122003} {\bibfield  {journal} {\bibinfo  {journal}
  {Theory of Probability \& Its Applications}\ }\textbf {\bibinfo {volume}
  {22}},\ \bibinfo {pages} {24} (\bibinfo {year} {1977})},\ \Eprint
  {https://arxiv.org/abs/https://doi.org/10.1137/1122003}
  {https://doi.org/10.1137/1122003} \BibitemShut {NoStop}%
\bibitem [{\citenamefont {Ellis}(1984)}]{ellis1984large}%
  \BibitemOpen
  \bibfield  {author} {\bibinfo {author} {\bibfnamefont {R.~S.}\ \bibnamefont
  {Ellis}},\ }\bibfield  {title} {\bibinfo {title} {Large deviations for a
  general class of random vectors},\ }\href
  {http://www.jstor.org/stable/2243592} {\bibfield  {journal} {\bibinfo
  {journal} {The Annals of Probability}\ }\textbf {\bibinfo {volume} {12}},\
  \bibinfo {pages} {1} (\bibinfo {year} {1984})}\BibitemShut {NoStop}%
\bibitem [{\citenamefont {Rockafellar}(1970)}]{rockafellar1970convex}%
  \BibitemOpen
  \bibfield  {author} {\bibinfo {author} {\bibfnamefont {R.~T.}\ \bibnamefont
  {Rockafellar}},\ }\href@noop {} {\emph {\bibinfo {title} {Convex
  analysis}}},\ Vol.~\bibinfo {volume} {18}\ (\bibinfo  {publisher} {Princeton
  university press},\ \bibinfo {year} {1970})\BibitemShut {NoStop}%
\bibitem [{\citenamefont {Lebowitz}\ and\ \citenamefont
  {Spohn}(1999)}]{lebowitz1999gallavotti}%
  \BibitemOpen
  \bibfield  {author} {\bibinfo {author} {\bibfnamefont {J.~L.}\ \bibnamefont
  {Lebowitz}}\ and\ \bibinfo {author} {\bibfnamefont {H.}~\bibnamefont
  {Spohn}},\ }\bibfield  {title} {\bibinfo {title} {A gallavotti--cohen-type
  symmetry in the large deviation functional for stochastic dynamics},\ }\href
  {https://doi.org/10.1023/A:1004589714161} {\bibfield  {journal} {\bibinfo
  {journal} {Journal of Statistical Physics}\ }\textbf {\bibinfo {volume}
  {95}},\ \bibinfo {pages} {333} (\bibinfo {year} {1999})}\BibitemShut
  {NoStop}%
\bibitem [{\citenamefont {Richardson}\ and\ \citenamefont
  {Walker}(1926)}]{richardson1926atmospheric}%
  \BibitemOpen
  \bibfield  {author} {\bibinfo {author} {\bibfnamefont {L.~F.}\ \bibnamefont
  {Richardson}}\ and\ \bibinfo {author} {\bibfnamefont {G.~T.}\ \bibnamefont
  {Walker}},\ }\bibfield  {title} {\bibinfo {title} {Atmospheric diffusion
  shown on a distance-neighbour graph},\ }\href
  {https://doi.org/10.1098/rspa.1926.0043} {\bibfield  {journal} {\bibinfo
  {journal} {Proceedings of the Royal Society of London. Series A, Containing
  Papers of a Mathematical and Physical Character}\ }\textbf {\bibinfo {volume}
  {110}},\ \bibinfo {pages} {709} (\bibinfo {year} {1926})}\BibitemShut
  {NoStop}%
\bibitem [{\citenamefont {Boffetta}\ and\ \citenamefont
  {Sokolov}(2002)}]{boffetta2002relative}%
  \BibitemOpen
  \bibfield  {author} {\bibinfo {author} {\bibfnamefont {G.}~\bibnamefont
  {Boffetta}}\ and\ \bibinfo {author} {\bibfnamefont {I.~M.}\ \bibnamefont
  {Sokolov}},\ }\bibfield  {title} {\bibinfo {title} {Relative dispersion in
  fully developed turbulence: The richardson's law and intermittency
  corrections},\ }\href {https://doi.org/10.1103/PhysRevLett.88.094501}
  {\bibfield  {journal} {\bibinfo  {journal} {Phys. Rev. Lett.}\ }\textbf
  {\bibinfo {volume} {88}},\ \bibinfo {pages} {094501} (\bibinfo {year}
  {2002})}\BibitemShut {NoStop}%
\bibitem [{\citenamefont {Cherstvy}\ \emph {et~al.}(2013)\citenamefont
  {Cherstvy}, \citenamefont {Chechkin},\ and\ \citenamefont
  {Metzler}}]{cherstvy2013}%
  \BibitemOpen
  \bibfield  {author} {\bibinfo {author} {\bibfnamefont {A.~G.}\ \bibnamefont
  {Cherstvy}}, \bibinfo {author} {\bibfnamefont {A.~V.}\ \bibnamefont
  {Chechkin}},\ and\ \bibinfo {author} {\bibfnamefont {R.}~\bibnamefont
  {Metzler}},\ }\bibfield  {title} {\bibinfo {title} {Anomalous diffusion and
  ergodicity breaking in heterogeneous diffusion processes},\ }\href
  {https://doi.org/10.1088/1367-2630/15/8/083039} {\bibfield  {journal}
  {\bibinfo  {journal} {New Journal of Physics}\ }\textbf {\bibinfo {volume}
  {15}},\ \bibinfo {pages} {083039} (\bibinfo {year} {2013})}\BibitemShut
  {NoStop}%
\bibitem [{\citenamefont {Stella}\ \emph {et~al.}()\citenamefont {Stella},
  \citenamefont {Chechkin},\ and\ \citenamefont
  {Teza}}]{stella2022inpreparation}%
  \BibitemOpen
  \bibfield  {author} {\bibinfo {author} {\bibfnamefont {A.~L.}\ \bibnamefont
  {Stella}}, \bibinfo {author} {\bibfnamefont {A.}~\bibnamefont {Chechkin}},\
  and\ \bibinfo {author} {\bibfnamefont {G.}~\bibnamefont {Teza}},\ }\bibinfo
  {title} {(to be published)}\BibitemShut {NoStop}%
\end{thebibliography}%

\end{document}